%Paper: hep-th/9405038
%From: etingof-pavel@MATH.YALE.EDU (Pavel Etingof)
%Date: Thu, 5 May 1994 18:58:52 -0400
%Date (revised): Sat, 21 May 1994 20:54:17 -0400

\input amstex
\magnification 1200
\documentstyle{amsppt}
\NoBlackBoxes
\NoRunningHeads
\def\g{\frak g}

\def\C{\Bbb C}

\def\d{\partial}

\def\Tr{\text{\rm Tr}}
\def\l{\lambda}

\topmatter
\title Integral formulas for wave functions
       of quantum many-body problems and representations
       of ${\frak gl}_n$
  \endtitle
\author {\rm {\bf  Pavel Etingof} \linebreak
	\vskip .1in
   Department of Mathematics\linebreak
   Yale University\linebreak
   New Haven, CT 06520, USA\linebreak
   e-mail: etingof\@math.yale.edu}
\endauthor
\endtopmatter

\centerline{April 30, 1994}
\centerline{hep-th 9405038}
\vskip .1in
\centerline{\bf Abstract}
\vskip .1in

We derive explicit integral formulas for eigenfunctions of quantum
integrals of the Calogero-Sutherland-Moser operator with trigonometric
interaction potential. In particular, we derive explicit
formulas for Jack's symmetric functions. To obtain such formulas,
we use the representation of these eigenfunctions
by means of traces of intertwining operators between certain modules
over the Lie algebra $\frak gl_n$,
and the realization of these
modules on functions of many variables.

\vskip .1in
\centerline{\bf Introduction}
\vskip .1in

Integrable many body problems were considered by F.Calogero \cite{C}
and B.Sutherland \cite{S} in early seventies.
Later more general Hamiltonian systems of this type were described (see
\cite{OP} for references).

An important progress in the theory of integrable
many-body problems was the discovery of their relationship to
the theory of Lie groups and symmetric spaces. In \cite{KKS},
it was found that the classical Calogero and Sutherland systems
can be obtained by Hamiltonian reduction from simpler looking
Hamiltonian systems (but having more degrees of freedom) arising
from Lie groups. Recently this approach was generalized
to infinite-dimensional groups \cite{GN1},\cite{GN2}, which allows to obtain
the elliptic deformation of the Sutherland system (see \cite{OP}),
and its $q$-deformation \cite{Ru}.

Quantum integrable systems of the Calogero-Sutherland type
first appeared in the theory of radial parts of Laplace's
operators on symmetric spaces (see \cite{H},\cite{W}). In the sixties
Harish-Chandra computed the radial part of the second order Laplace's operator
acting on generalized spherical functions on a homogeneous space
for a semisimple Lie group $G$ (see \cite{W}). One can show that
Sutherland's quantum Hamiltonian can be obtained as a special case
of such radial part. This implies complete integrability of
Sutherland`s Hamiltonian: its integrals are just the radial parts
of the higher Laplacians, obtained from the center of the universal
enveloping algebra of Lie$G$ (which happens to be ${\frak gl}_n$ in
this case). Comparing this to the classical situation, one can say
that the procedure of computation of the radial part
is a quantum counterpart of Hamiltonian reduction.

A proof of integrability of the Sutherland Hamiltonian
and a construction of its wave functions based on these ideas
(but using a more algebraic method) is given in \cite{E}.
The approach of \cite{E} is to represent the wave functions
as weighted traces of intertwining operators between certain
representations of $G$. The goal of the present paper is to use
this approach to give an explicit integral formula for wave functions.

The paper is organized as follows. In Section 1 we describe
the construction of the wave functions as traces of intertwiners.
In Section 2 we introduce the realization of representations of ${\frak gl}_n$
by differential operators, and use this realization to represent
intertwiners as functions of many variables. This allows
to write down an integral formula for the trace of such intertwiner,
which is our main result (Theorem 2.2). As a corollary, we get an explicit
integral representation of Jack's symmetric functions. It follows from
Theorem 2.2 and the results of \cite{EK1},\cite{EK2}.

\vskip .1in
\centerline{\bf 1. Trace representation of wave functions}
\vskip .1in

The Hamiltonian of the quantum $n$-body problem on the line is
$$
H=\sum_{i=1}^n\frac{\d^2}{\d x_i^2}-C\sum_{i<j}U(x_i-x_j),\tag 1.1
$$
where $U$ is some potential function, and $C$ is a constant.
We will consider the case $U(x)=\frac{1}{2\text{sinh}^2(x/2)}$.
This case was first considered by Sutherland \cite{S}; so we call
the operator $H$ the Sutherland operator. It is a trigonometric
deformation of the Calogero operator \cite{C}, which is given by (1.1)
with $U(x)=x^{-2}/2$.

It is known (cf. \cite{OP}) that the Sutherland operator defines
a completely integrable quantum system. This means, it has $n$
pairwise commutative and algebraically independent quantum integrals
$$
L_1=\sum_{i}\frac{\d}{\d x_i},\ L_2=H+\text{const},
\ L_m=\sum_{i}\frac{\d^m}{\d
x_i^m}+\text{lower order terms}, 3\le m\le n\tag 1.2
$$
(the choice of $L_m$ is not unique). Wave functions of this system
are, by definition, joint eigenfunctions of $L_1,...,L_n$, i.e.
solutions of the differential equations
$$
L_i\psi=\Lambda_i\psi, \ 1\le i\le n,\tag 1.3
$$
where $\Lambda_i$ are some complex numbers. The space of such
functions is $n!$-dimensional for every set of $\{\Lambda_i\}$.

In this paper we present an explicit integral formula for wave
functions. It is given by Theorem 2.2.

The first step in deriving this formula is a
representation of wave functions as traces of intertwining
operators between certain representations of the Lie algebra
${\frak gl}_n$.

Let $\g$ denote the Lie algebra ${\frak gl}_n$, $\frak h=\C^n$ denote
the Cartan subalgebra of diagonal matrices, $M_{\l}$ denote
the Verma module over $\g$ with highest weight $\l$, and
let $U_k$ ($k\in\C$) be the space of functions
$$
U_{k}=\{(z_1\ldots z_n)^{k-1} p(z), p(z)\in \C[z_1^{\pm 1},\ldots,
z_n^{\pm 1}], \text{deg }p=0\}\tag 1.4
$$
with the action of $\g$ given by
$$
E_{ij}\to z_i\frac{\d}{\d z_j}-(k-1)\delta_{ij},\tag 1.5
$$
where $E_{ij}$ are the elementary matrices: $(E_{ij})_{lm}=\delta_{il}
\delta_{jm}$.

\proclaim{Lemma 1.1} If $M_{\l}$ is irreducible then
there exists a unique up to a factor $\g$-intertwining operator
$\Phi_{\l}:M_{\l}\to M_{\l}\hat\otimes U_k$, where
$\hat\otimes$ denotes a completed tensor product.
\endproclaim

Consider the function
$$
\Psi(x,\l)=\Tr|_{M_{\l}}(\Phi_{\l}e^x),\ x\in \frak h.\tag 1.6
$$
This is a series defining an analytic function in the region
$\text{Re}x_1>...>\text{Re}x_n$, where $x_i=<x,h_i>$, $h_i=E_{ii}$.
This function takes values in the zero-weight subspace $U_k[0]$ of
$U_k$, which is one-dimensional. Thus, we will regard it as a scalar function.

\proclaim{Theorem 1.2}\cite{E}
 Let $Z$ be any element of the center of $U(\g)$. Then there exists
a unique differential operator $D_Z$ on functions of $n$ variables
$x_1,...,x_n$ dependent on $k$ but not on $\l$ such that for a generic $\l$
$$
\Tr|_{M_{\l}}(\Phi_{\l}Ze^x)=D_Z \Tr|_{M_{\l}}(\Phi_{\l}e^x).\tag 1.7
$$
\endproclaim

Consider the free generators of the center of $U(\g)$:
$Z_m=\sum_{i_1,...,i_m=1}^nE_{i_1i_2}\dots
E_{i_{m-1}i_m}E_{i_mi_1}$, $1\le m\le n$.
Set $\hat L_m=D_{Z_m}$.

Also, introduce the Weyl denominator
$$
\phi(x)=2^{n(n-1)/2}\prod_{1\le i<j\le
n}\text{sinh}(\frac{x_i-x_j}{2}).
\tag 1.8
$$

Define the constant $C$ in (1.1) by
$$
C=k(k-1).\tag 1.9
$$

\proclaim{Theorem 1.3}\cite{E} The operators
$\hat L_1,...,\hat L_m$ are pairwise commutative
and simultaneously conjugate to quantum integrals of the Sutherland
operator: if one defines $L_m$ by the formula
$L_mf=\phi\hat L_m\left(\frac{f}{\phi}\right)$
then $L_2=H-<\rho,\rho>$, $\rho=(\frac{n-1}{2},...,\frac{1-n}{2})$,
and $L_m$ are quantum integrals of the Sutherland operator
defined by (1.2).
\endproclaim

 From now on we will assume that the choice of $L_m$ is made
according to Theorem 1.3.

Let $P_i(\l)$ be the scalar by which $Z_i$ acts in $M_{\l}$
(it is a symmetric polynomial of $\l+\rho$). Consider the function
$$
\psi(x,\l)=\phi(x)\Psi(x,\l).\tag 1.10
$$

Let us say that $\l$ is generic if
$\l_i-\l_j+i-j$ is not a positive integer when $i>j$.
A weight $\l$ is generic in this sense iff the corresponding
Verma module $M_{\l}$ is irreducible.

\proclaim{Corollary 1.4}\cite{E} (i) The function $\psi$ is an
eigenfunction of $L_1,...,L_n$ ( a wave function) for a generic $\l$:
$$
L_i\psi(x,\l)=P_i(\l)\psi(x,\l).\tag 1.11
$$

(ii) If the weights $w(\l+\rho)-\rho$ are generic
for any $w\in S_n$ ($S_n$ is the symmetric group) then
the functions $\psi(x,w(\l+\rho)-\rho), w\in S_n$ form a basis in the space of
solutions of (1.11).
\endproclaim

This corollary shows that in order to compute joint eigenfunctions
of $L_i$ for generic eigenvalues, it is enough to calculate
the trace function $\Psi(x,\l)$. In the subsequent sections
we will do it using the Borel-Weil realization
(``bosonization'') of the Lie algebra $\frak{gl}_n$.
\vskip .1in

\centerline{\bf 2. The Borel-Weil realization of ${\frak gl}_n$ and
integral formulas}
\vskip .1in

By the Borel-Weil realization we mean the realization of an irreducible
Verma module $M_{\l}$ over ${\frak gl}_n$ in the space of regular
functions on the big cell of the flag variety; the action of the Lie
algebra in this space is given by first order differential operators.

Let $B^+$ be the Borel subgroup in $G=GL_n$ consisting of upper
triangular matrices. Let $C_0$ be the big cell of the Bruhat
decomposition of $G$. Let $\tilde B^+$, $\tilde C_0$ be the universal
covers of $B^+$, $C_0$. We have a left action of $\g$ on $\tilde C_0$
which commutes with the (free) right action of $\tilde B^+$.
 Consider the induced action of $\g$
on the space $M_{\l}^c$ of regular functions
$f:\tilde C_0\to\C$ satisfying the condition
$f(xb)=\chi_{\l}(b)f(x)$, $b\in \tilde B^+$, where $\chi_\l:\tilde
B^+\to \C^*$ is the character of $\tilde B^+$ obtained by extention of
$\l\in \frak h^*$. Clearly, the space $M_{\l}^c$ can be realized
as the space of regular functions on the big Schubert cell
$F_0=C_0/B^+$. This cell can be naturally realized as the set of all
strictly upper triangular matrices. Thus, we have a representation of
$\g$ in the space of functions of strictly upper triangular matrices.

Explicitly, this representation looks as follows.

Set $M_{\l}^c=\C[\{y_{ij}\}]$ (as a vector space),
where $\{y_{ij},1\le i<j\le n\}$ is a collection of independent variables.
Define the action of $\g$ on $M_{\l}^c$ as follows:
$$
\gather
E_{ii}-E_{i+1,i+1}=2y_{ii+1}\frac{\d}{\d
y_{ii+1}}+\sum_{j=1}^{i-1}(y_{ji+1}\frac{\d}{\d y_{ji+1}}-
y_{ji}\frac{\d}{\d y_{ji}})\\
+\sum_{j=i+1}^{n-1}
(y_{ij+1}\frac{\d}{\d y_{ij+1}}-
y_{i+1,j+1}\frac{\d}{\d y_{i+1,j+1}})+\lambda_i-\l_{i+1}, 1\le i\le n-1,\\
E_{i+1i}=\frac{\d}{\d
y_{ii+1}}+\sum_{j=i+1}^{n-1}y_{i+1,j+1}\frac{\d}{\d y_{i,j+1}}, 1\le
i\le n-1,\\
E_{ii+1}=\sum_{j=1}^{i-1}y_{ii+1}y_{ji}\frac{\d}{\d y_{ji}}-
\sum_{j=1}^i y_{ii+1}y_{ji+1}\frac{\d}{\d
y_{ji+1}}-(\l_i-\l_{i+1})y_{ii+1}\\
+\sum_{j=i+1}^{n-1}y_{ij+1}\frac{\d}{\d y_{i+1,j+1}}-
\sum_{j=1}^{i-1}y_{ji+1}\frac{\d }{\d y_{ji}},\ 1\le i\le n-1,
\\
\sum_{i}E_{ii}=\sum_{i}\lambda_i.\tag 2.1
\endgather
$$
(These formulas are borrowed from \cite{FF})
{}From these formulas one can deduce by recursion the
explicit form of the action of $E_{ij}$ in $M_\l^c$ for arbitrary $i,j$.
One can do so by using (2.1) and the relations
$$
E_{ij}=\cases [E_{ii+1},...[E_{j-2,j-1},E_{j-1,j}]..],& i<j\\
 [E_{ii-1},...[E_{j+2,j+1},E_{j+1,j}]..],& i>j\endcases\tag 2.2
$$
The obtained formulas will have the general form
$$
E_{ij}=\sum_{l<m}A_{ijlm}(y,\l)\frac{\d}{\d y_{lm}}+b_{ij}(y,\l),\tag 2.3
$$
where $A_{ijkl},b_{ij}$ are polynomials determined from (2.1), (2.2).

The module $M_{\l}^c$ is isomorphic to the contragredient Verma
module, but if $\l$ is generic, it is also isomorphic to the usual
Verma module $M_{\l}$. Further we assume that $\l$ is generic.

Let us now present a functional realization of the restricted
dual of a Verma module, $M_{\l}^*$. We will realize
it in the space $\C[\{t_{ij}^{-1}\}]$, where $t_{ij}$, $i<j$, are
independent variables. Geometrically this space is the space
of regular functions on the image of the big cell $F_0$ under the
action of the longest element in the Weyl group.
Define the pairing between $M_{\l}$
 and $M_{\l}^*$ by the formula
$$
(f,g)=(2\pi \sqrt{-1})^{-n(n-1)/2}\int_{|y_{ij}|=1}f(y)g(y)\frac{dy}{y},\ f\in
M_\l,\ g\in M_\l^*,\tag 2.4
$$
where by definition $\frac{dy}{y}=
\wedge_{i=1}^n\wedge_{j=i+1}^n\frac{dy_{ij}}{y_{ij}}$.

We would like this pairing to be $\g$-invariant, which uniquely
determines the action of $\g$ in $M_{\l}^*$:
$$
E_{ij}=\sum_{l<m}A_{ijlm}(t,\l)\frac{\d}{\d t_{lm}}+b_{ij}^*(t,\l),
$$
where
$$
b_{ij}^*(t,\l)=\sum_{l<m}(\frac{\d A_{ijlm}}{\d t_{lm}}-
\frac{A_{ijlm}}{t_{lm}})-b_{ij}(t,\l).\tag 2.5
$$

Now we can give a functional relaization of
the intertwiner $\Phi_{\l}$ which was
defined in Section 1. We can regard $\Phi_{\l}$ as an element
of the space $M_{\l}\hat\otimes M_{\l}^*\hat\otimes U_{k}$, i.e. as a
function of $y_{ij}$, $t_{ij}$, $z_i$: $\Phi_\l=\Phi_\l
(y,t,z)$.
This function is invariant under the action of the Cartan subalgebra,
which implies that it can be written as
$$
\Phi_{\l}(y,t,z)=(z_1...z_n)^{k-1}\Theta_{\l}(\xi,\eta),\tag 2.6
$$
where $\xi_{ij}$, $\eta_{ij}$ are the new variables:
$\xi_{ij}=y_{ij}z_j/z_i$, $\eta_{ij}=t_{ij}z_j/z_i$.
 Also, (2.6) satisfies the system of differential equations
$E_{ij}\Phi_{\l}=0$,
which can be rewritten in the new coordinates as follows:
$$
\gather
\sum_{l<m}A_{ijlm}(\xi,\l)\frac{\d\Theta_\l}{\d \xi_{lm}}+
\sum_{l<m}A_{ijlm}(\eta,\l)\frac{\d\Theta_\l}{\d \eta_{lm}}+\\
\sum_{p<j}(\xi_{pj}\frac{\d\Theta_{\l}}{\d \xi_{pj}}-
\eta_{pj}\frac{\d\Theta_{\l}}{\d \eta_{pj}})-
\sum_{p>j}(\xi_{jp}\frac{\d\Theta_{\l}}{\d \xi_{jp}}-
\eta_{jp}\frac{\d\Theta_{\l}}{\d \eta_{jp}})+\\
(b_{ij}(\xi,\l)+b_{ij}^*(\eta,\l)+(k-1))\Theta_{\l}=0, 1\le i,j\le
n, i\ne j\tag 2.7
\endgather
$$

System (2.7) can be regarded as a system of $n^2-n$ nonhomogeneous
linear equations with respect to the $n^2-n$ variables
$\frac{\d\Theta_\l}{\d\xi_{ij}},\frac{\d\Theta_\l}{\d\eta_{ij}}$.
Denote the matrix of this system by $N(\xi,\eta)$. It is a square
matrix.

\proclaim{Lemma 2.1} The matrix $N(\xi,\eta)$ is invertible
over the field of rational functions of $\xi,\eta$. The inverse
matrix is regular in the neighborhood of the point $\xi=0$, $\eta=\infty$.
\endproclaim

\demo{Proof} Let us show that the vector fields by
which the $n^2-n$ elements $E_{ij},i\ne j$ act on the space
$F\times F$, where $F$ is the flag variety, are linearly independent
at the point $(B^+,B^-)$, where $B^+,B^-$ are the Borel subgroups
in $G=GL_n(\C)$ consisting of upper (respectively lower) triangular
matrices. Indeed, the stabilizer of $B^{\pm}$ is $B^{\pm}$ itself,
so the stabilizer of both $B^+,B^-$ is $B^+\cap B^-=H$ -- the maximal torus
consisting of diagonal matrices. Thus, the
orbit of $(B^+,B^-)$ is isomorphic to $G/H$, and its tangent space
at $(B^+,B^-)$ is $\g/\frak h$. The elements $E_{ij}$ project onto a
basis of $\g/\frak h$, so they produce linearly independent vectors
on the tangent space.
\qed\enddemo

Lemma 2.1 implies that system of differential equations (2.7) can be
written in the form
$$
\gather
\frac{\d \Theta_\l}{\d \xi_{ij}}=\alpha_{ij}(\xi,\eta)\Theta_\l,\\
\frac{\d \Theta_\l}{\d \eta_{ij}}=\beta_{ij}(\xi,\eta)\Theta_\l,\tag 2.8
\endgather
$$
where $\alpha_{ij},\beta_{ij}$ are rational functions of $\xi, \eta$
found from (2.7) (all functions from (2.7) implicitly depend on $k$).
This determines $\Theta_\l$ uniquely up to a
constant. If we normalize $\Theta_{\l}$ by the condition
$\Theta_{\l}(0,\infty)=1$ then $\Theta_{\l}$ is given by the formula
$$
\Theta_\l(\xi^0,\eta^0)=\text{exp}\biggl(\int_{(0,\infty)}^{(\xi^0,\eta^0)}
\sum_{i<j}(\alpha_{ij}(\xi,\eta)d\xi_{ij}+\beta_{ij}(\xi,\eta)d\eta_{ij})
\biggr).\tag 2.9
$$
 It follows from the above that this function
is regular in the neighborhood of $(0,\infty)$.

Now we can deduce integral formulas for $\Psi(x,\l)$.
The idea is that $\Psi$ is obtained from $\Phi_{\l}\in M_{\l}\otimes
M_{\l}^*\otimes U_k$ by contracting $M_{\l}$ with $M_{\l}^*$.
This contraction, in the functional realization, is expressed by a
contour integral (see (2.4)). Therefore, from (1.6) and (2.6) we get
$$
\Psi(x,\l)=
(2\pi\sqrt{-1})^{-n(n-1)/2}e^{<\l,x>}
\int_{|y_{ij}|=1}\Theta_{\l}(\{y_{ij}\},\{e^{x_i-x_j}y_{ij}\})
\frac{dy}{y}.\tag 2.10
$$

Thus, we have proved the following theorem, which is our main result.

\proclaim{Theorem 2.2} For generic values of $\Lambda_i$ in (1.3)
there exists $\mu\in\C^n$ such that the functions
$$
\gather
\psi(x,\l)=\\
\phi(x)e^{<x,\l>}
\int_{|y_{ij}|=1}\text{exp}\biggl(\int_{(0,\infty)}^{(\{y_{ij}\},
\{e^{x_i-x_j}y_{ij}\})}
\sum_{i<j}(\alpha_{ij}(\xi,\eta,\l,k)d\xi_{ij}+\beta_{ij}(\xi,\eta,\l,k)d\eta_{ij})
\biggr)\frac{dy}{y},\\
\l=w(\mu+\rho)-\rho, w\in S_n,\tag 2.11\endgather
$$
where $\alpha_{ij}$, $\beta_{ij}$ are the rational functions defined
above\footnote{In (2.11) we explicitly specify the dependence of
$\alpha_{ij}$, $\beta_{ij}$ on $\l,k$, which we have omitted so far.},
$C=k(k-1)$, and $\phi$ is given by (1.8),
form a basis of the space of solutions of system (1.3).
The weight $\mu$ is found from the equations $P_i(\mu)=\Lambda_i$,
$1\le i\le n$, where $P_i(\l)$ is the scalar by which the
central element $Z_m$ acts in the Verma module $M_\l$.
\endproclaim

{\bf Remark. } It is known that if $n=2$ then eigenfunctions of
$L_1,L_2$ express via Gauss hypergeometric function. It is easy to
check that for $n=2$ formula (2.11) becomes (after some
transformations) a special case of the standard integral formula for the Gauss
hypergeometric function.

Now we can apply Theorem 2.2 to derivation of integral formulas
for Jack's symmetric functions (see \cite{M}).
In \cite{EK1},\cite{EK2} it was shown that the Jack's symmetric
functions can be obtained as follows:
$$
J_\l^k(x)=\phi(x)^{1-k}\Tr|_{M_{\l+(k-1)\rho}}(\Phi_{\l+(k-1)\rho}e^x)
\tag 2.12
$$
This formula combined with Theorem 2.2 gives the desired integral
representation of Jack's functions.

{\bf Remark. } For integer values of $k$, the representation $U_k$
contains a finite dimensional subrepresentation or quotient
representation. Therefore, in this case common eigenfunctions of $\hat L_i$
can be interpreted as generalized spherical functions on the symmetric
space $K\times K/K_{diag}$, where $K=SU(n)$, and $K_{diag}$ is the
diagonal in $K\times K$. Therefore,
it is possible to write a Harish-Chandra type integral formula
for these eigenfunctions, as described in \cite{W}, where integration
is carried out over $K$. It is an interesting question what is the
connection of this formula with the one given by Theorem 2.2.

\vskip .1in

{\bf Acknowledgements. } The author would like to thank
I.Cherednik, I.Frenkel, A.Kirillov Jr., and A.Varchenko, and G.Zuckerman
for discussions.

\end
\Refs
\ref\by [Ca] Calogero, F.\paper Solution of the one-dimensional n-body
problem with quadratic and/or inversely quadratic pair potentials
\jour J. Math. Phys. \vol 12\pages 419-436\yr 1971\endref

\ref\by [E] Etingof, P.I.\paper Quantum integrable systems and
representations of Lie algebras, hep-th 9311132\jour submitted to
Journal of Mathematical Physics\yr 1993\endref

\ref\by [EK1] Etingof, P.I. and Kirillov, A.A., Jr\paper A unified
representation-theoretic approach to special functions, hep-th 9312101
\jour Functional Anal. and its Applic.\vol 28\issue 1
 \yr 1994\endref

\ref\by [EK2] Etingof, P.I. and Kirillov, A.A., Jr\paper Macdonald's
polynomials and representations of quantum groups \jour to appear in
Math. Res. Let.\yr 1994\endref

\ref\by [FF] Feigin, B.L. and Frenkel, E.V.\paper
Representations of affine Kac-Moody algebras and bosonization,
\book V.Knizhnik Memorial Volume, eds L.Brink, D.Freidan, A.M.Polyakov
\publ World Scientific \publaddr Singapore\yr 1990 \pages 271-316
\endref

\ref\by [GN1] Gorsky, A., and Nekrasov, N. \paper Relativistic
Calogero-Moser Model as Gauged WZW theory\jour hep-th 9401017
\yr 1993\endref

\ref\by [GN2] Gorsky, A., and Nekrasov, N. \paper
Elliptic Calogero-Moser system from two-dimensional current algebra
\jour hep-th 9401021\yr 1994\endref

\ref\by [H] Helgason, S. \book Differential geometry, Lie groups, and
symmetric spaces\publ Academic Press\publaddr New York\yr 1978\endref

\ref\by [KKS] D.Kazhdan, B.Kostant, and S.Sternberg, \paper
Hamiltonian group actions and dynamical systems of Calogero type
\jour Comm. Pure and Appl.Math. \vol 31\pages 481-507\yr 1978\endref

\ref\by [M] Macdonald, I.G. \paper A new class of symmetric
functions\jour Publ. I.R.M.A. Strasbourg, 372/S-20, Actes 20
S\'eminaire Lotharingien\pages 131-171\yr 1988\endref

\ref\by [OP] Olshanetsky, M.A., and Perelomov, A.M.\paper Quantum
integrable systems related to Lie algebras\jour Phys. Rep. \vol
94\pages 313-404\yr 1983\endref

\ref\by [Ru] Ruijsenaars, S.N.M. \jour Comm. Math. Phys. \vol
110\pages 191-213\yr 1987\endref

\ref\by [S] Sutherland, B.\paper Exact results for a quantum many-body
problem in one dimension
\jour Phys. Rev. A5\pages 1372-1376\yr 1972\endref

\ref\by [W] Warner, G. \book Harmonic analysis on semi-simple Lie
groups II\publ Springer-Verlag\yr 1972\endref
